\documentclass[%
  aip,%
  ajp,%
  reprint,
  superscriptaddress,%
  longbibliography
]{revtex4-2}

\usepackage{graphicx}
\usepackage{amsmath,amssymb}
\usepackage{hyperref}

\begin{document}

\title{World without Viscosity}

\author{Mohammad-Reza Alam}
\affiliation{Department of Mechanical Engineering, University of California, Berkeley, CA 94563, USA}


\date{\today}

\begin{abstract}
Viscosity - the internal friction of fluids - is among the most consequential yet underappreciated properties in physics. This paper explores what would happen if viscosity vanished from all fluids while other material properties remained unchanged. The consequences are catastrophic and universal. Aircraft cannot generate lift because circulation around wings requires viscous action. Rotating machinery seizes without lubricating fluid films. Cardiovascular systems lose the resistance necessary for pressure regulation. Rivers become violent torrents, aquifers drain in hours, and storms persist indefinitely without frictional dissipation. The pedagogical value lies in illuminating viscosity's role providing resistance, damping, and control across all scales - from cellular interiors to planetary atmospheres. Evolution, engineering, and climate have exploited viscous dissipation for billions of years; its absence would render complex life impossible and Earth uninhabitable. By imagining a world without viscosity, we better understand the viscous world we inhabit.
\end{abstract}

\maketitle

\section{Introduction}
In 1884, the English schoolmaster Edwin Abbott published \emph{Flatland}\cite{abbott1884flatland}, a slim novella that would become one of the most enduring thought experiments in scientific literature. The premise was deceptively simple: what if intelligent beings lived in a two-dimensional world? Abbott's inhabitants - squares, triangles, and circles - could only perceive each other as lines. Through this constrained universe, Abbott invited readers to question assumptions so fundamental that we rarely notice them. The book was not merely about geometry; it was about the invisible architecture of experience.

\emph{Flatland} belongs to a small but distinguished tradition of speculative works that ask: what if a basic feature of our physical world were different? Greg Egan's \emph{Orthogonal} trilogy imagines a universe with altered spacetime geometry\cite{egan2011clockworkrocket,egan2012eternalflame,egan2013arrowsoftime}. Hal Clement's \emph{Mission of Gravity} explores a planet where surface gravity varies from $3g$ to $700g$\cite{clement1954missionofgravity}. Robert Forward's \emph{Dragon's Egg} depicts life on a neutron star under $67\times 10^{9}\,g$\cite{forward1980dragonsegg}. Each of these works uses rigorous extrapolation to illuminate how deeply physics shapes everyday existence.

This article attempts a similar exercise. We will imagine a world in which one particular physical property, namely viscosity, has been switched off. We assume that dry friction between solid surfaces still operates normally, but the internal friction of fluids, the resistance that arises when layers of liquid or gas slide past one another, is gone. What would such a world look like? Before we explore the consequences, we must first appreciate what viscosity is and why it deserves this treatment.

Viscosity is, in essence, a fluid's resistance to flow. Pour water and pour honey; the difference we feel is viscosity. More precisely, viscosity quantifies the internal friction that arises when adjacent layers of fluid move at different speeds. When we stir a cup of tea, the liquid near the spoon moves faster than the liquid near the cup's wall. Viscosity is what transfers momentum between these layers, what makes the whole fluid eventually spin together, and what dissipates our stirring energy into heat.

The range of viscosities in nature is staggering. Water at room temperature has a dynamic viscosity of about $\mu \approx 1\times 10^{-3}\,\mathrm{Pa\cdot s}$ (one millipascal-second, or one centipoise in older units). Air is roughly $50$ times less viscous, at about $\mu \approx 1.8\times 10^{-5}\,\mathrm{Pa\cdot s}$. Honey, depending on temperature and type, ranges from $2$ to $10\,\mathrm{Pa\cdot s}$, thousands of times more viscous than water. At the extremes, window glass behaves as a fluid with viscosity around $10^{12}\,\mathrm{Pa\cdot s}$, while the Earth's mantle creeps along at approximately $10^{21}\,\mathrm{Pa\cdot s}$. This spans nearly $27$ orders of magnitude, from air to mantle rock.

Viscosity governs phenomena at every scale of the physical world. In our body, blood viscosity (about $3$ to $4$ times that of water) determines how hard our heart must work: ``a world without viscosity would be a world without the human heart''\cite{zamir2000hemodynamics}. A $20\%$ increase in blood viscosity measurably raises cardiovascular strain. In the atmosphere, the viscosity of air enables the boundary layers that let airplanes fly and causes the drag that slows raindrops to terminal velocity. In geology, mantle viscosity controls the rate of continental drift and the timescale of post-glacial rebound: The ground beneath Scandinavia is still rising, centimeters per century, because the mantle is slowly relaxing from the weight of ice sheets that melted $10{,}000$ years ago.

\begin{figure}[h]
    \centering
    \includegraphics[width=0.8\columnwidth]{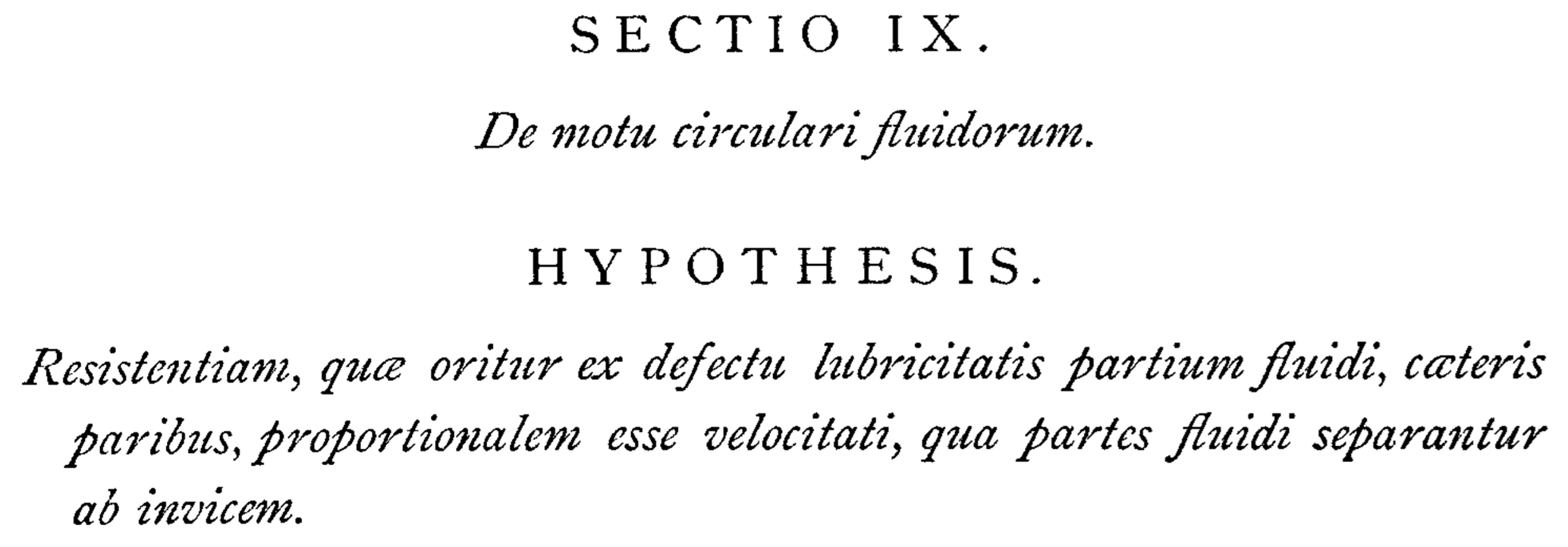}
    \includegraphics[width=0.9\columnwidth]{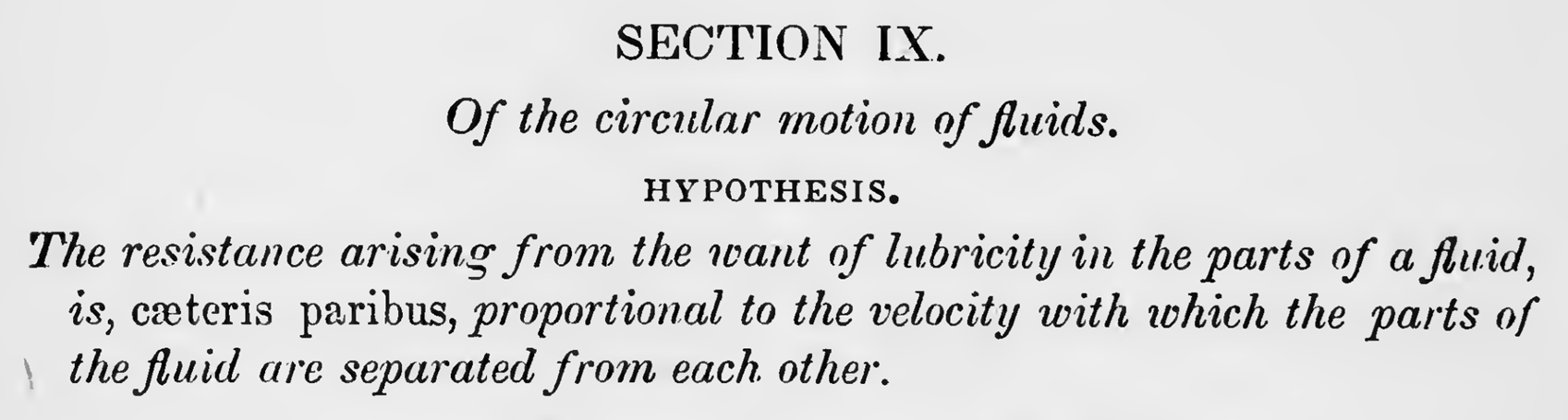}
    \caption{Newton's law of viscosity as it originally appeared in the Principia\cite{newton1687principia}, Book II, Section IX, P. 374 original Latin text (top figure), and p. 370, Motte translation 1729 (bottom figure). The Hypothesis states that shear stress is proportional to velocity gradient, expressed as resistance from ``want of lubricity" (Newton's word for what we call today \textit{viscosity}) being proportional to the rate at which fluid layers separate. In modern notation: $\tau = \mu \, du/dy$.}
    \label{fig:principia}
\end{figure}

Despite this ubiquity, viscosity remains underappreciated in public understanding of physics. Ask a student to name Newton's contributions, and you will hear about gravity, calculus, and the laws of motion. You are unlikely to hear about Newton's law of viscosity, which he formulated in the \emph{Principia}\cite{newton1687principia} alongside his more famous laws (Figure \ref{fig:principia}). Viscosity lacks the dramatic narrative of gravity or the conceptual elegance of energy conservation. It is a property we experience constantly, every breath, every heartbeat, every sip of coffee, yet rarely name. Perhaps its low profile stems from difficulty imagining its absence: we can picture weightlessness or frictionless planes, but a world without viscosity lacks visual vocabulary.

The theoretical study of inviscid flow has a long and paradoxical history. In 1752, Jean le Rond d'Alembert proved mathematically that an object moving through an ideal fluid with zero viscosity should experience zero drag; a result so contrary to experience that it became known as d'Alembert's paradox\cite{dalembert1752resistance}. As he wrote, ``I do not see then, I admit, how one can explain the resistance of fluids by the theory in a satisfactory manner.'' The resolution came only in 1904, when Ludwig Prandtl introduced the boundary layer concept\cite{prandtl1904boundarylayer}, showing that viscosity, however small, fundamentally changes flow behavior near surfaces. Feynman, in his celebrated lectures, referred to inviscid fluid as ``dry water''\cite{feynman1964lecturesv2}: a substance that obeys the Euler equations but bears little resemblance to the wet reality we experience (The phrase is sometimes attributed to John von Neumann, who reportedly used it to criticize inviscid fluid theory as unrealistic).

Questions about a world without viscosity - or any counterfactual physics - are not merely entertaining. They are pedagogically powerful. By removing a property we take for granted, we are forced to see its fingerprints everywhere. In our case of the world without viscosity, the thought experiment reveals viscosity's role in stabilizing flows, damping oscillations, enabling locomotion, and permitting the very existence of weather patterns. It shows why certain things that seem simple, (e.g. pouring a liquid, flying through air, or pumping blood) are in fact delicately dependent on this single material property.

Our thought experiment is not entirely without physical precedent. Nature does provide a window into zero-viscosity behavior through superfluidity: a quantum mechanical state first observed in liquid helium-4 ($^{4}\mathrm{He}$) by Pyotr Kapitsa and independently by John F. Allen and Don Misener in 1937\cite{kapitsa1938helium,allenmisener1938helium}. Below the lambda point at $T_{\lambda}=2.17\,\mathrm{K}$, helium undergoes a phase transition into a state where a fraction of the liquid behaves as a true zero-viscosity superfluid. This superfluid component can flow through capillaries less than $1\,\mu\mathrm{m}$ in diameter without resistance, creep up container walls as a thin Rollin film, and maintain circulation indefinitely without dissipation. More recently, researchers have achieved room-temperature superfluidity in polariton condensates which are hybrid light-matter quasiparticles confined in organic microcavities: In 2017, Lerario and colleagues demonstrated\cite{lerario2017roomtempsuperfluidity} superfluid flow at ambient conditions using exciton-polaritons in a Fabry-Perot cavity, opening possibilities for table-top studies of quantum hydrodynamics. While these systems involve quasiparticles rather than bulk fluids, they offer glimpses of what zero-viscosity behavior actually looks like\footnote{Note that quantum superfluids don't behave exactly like classical inviscid fluids. Quantized vortices, for instance, are a quantum phenomenon without classical analog.}.

In the sections that follow, I will systematically explore what a zero-viscosity world would look like. We will examine locomotion, sound, weather, biology, and everyday objects. The goal is not to predict a realistic alternative universe, but to use the contrast to illuminate our own. By imagining a world where this single property is absent, we reveal just how profoundly viscosity shapes the physics of daily life: from the motion of air in our lungs to the stability of ocean currents. In the spirit of Abbott's \emph{Flatland}, we will use an impossible world to better understand the possible one we inhabit.

\section{Fundamental Physics of Inviscid Flow}

We now turn to the mathematical foundations. What equations govern this strange world, and what consequences follow?

\subsection{Governing equations}

The incompressible Euler equations govern inviscid fluid motion. Conservation of momentum reads
\begin{equation}\label{euler}
\frac{\partial \mathbf{u}}{\partial t} + (\mathbf{u} \cdot \nabla)\mathbf{u} = -\frac{1}{\rho}\nabla p + \mathbf{g},
\end{equation}
where $\mathbf{u}$ is velocity, $t$ is time, $\rho$ is density, $p$ is pressure, and $\mathbf{g}$ is body force per unit mass. The left-hand side captures fluid acceleration (local plus convective); the right-hand side represents pressure and body forces. Conspicuously absent is the viscous term $\nu\nabla^2\mathbf{u}$ from Navier-Stokes. Without it, momentum cannot diffuse - it can only be carried bodily by the flow. Conservation of momentum, together with the conservation of mass,
\begin{equation}\label{consmass}
\nabla \cdot \mathbf{u} = 0,
\end{equation}
forms the Euler's equation. Taking the curl of the momentum equation \eqref{euler} and with the use of continuity \eqref{consmass} we arrive at the vorticity equation
\begin{equation}
\frac{D\boldsymbol{\omega}}{Dt} = (\boldsymbol{\omega} \cdot \nabla)\mathbf{u} + \frac{1}{\rho^2}\nabla\rho \times \nabla p + \nabla \times \mathbf{g},
\end{equation}
where $\boldsymbol{\omega} = \nabla \times \mathbf{u}$ is vorticity and $D/Dt$ is the material derivative. Each right-hand term represents a distinct mechanism of vortex generation:
\begin{itemize}
\item $(\boldsymbol{\omega} \cdot \nabla)\mathbf{u}$ is vortex stretching/tilting. It modifies/amplifies existing vorticity but cannot create it.
\item $\frac{1}{\rho^2}\nabla\rho \times \nabla p$ is the baroclinic torque. It generates vorticity when density and pressure gradients are misaligned, e.g. when an interface of oil-water is perturbed.
\item $\nabla \times \mathbf{g}$: non-conservative body forces (e.g. Coriolis or Lorentz). It can generate vorticity directly.
\end{itemize}

In the absence of initial vorticity ($\boldsymbol{\omega}=0$), the stretching/tilting term contributes nothing - it can amplify or reorient existing vortices, but it cannot conjure vorticity from an irrotational flow. Non-conservative body forces, the third term, are relatively rare guests in everyday fluid mechanics. They appear primarily in two settings: at planetary scales, where the Coriolis force orchestrates the grand circulation of atmospheres and oceans, and in magnetohydrodynamics, where the Lorentz force acts on conducting fluids - a phenomenon more at home in industrial processes than in daily experience.

Yet an inviscid world is not necessarily quiescent. The baroclinic term is the workhorse of vorticity generation. Whenever density and pressure gradients are misaligned, vorticity is born. This mechanism dominates in geophysical flows. Sunlight warming land more than the adjacent ocean creates horizontal density gradients that cross vertical pressure gradients - and a sea breeze spins up in response. On larger scales, baroclinic instability fuels the midlatitude weather systems.

In barotropic flow with conservative body forces, Kelvin's circulation theorem states
\begin{equation}
\frac{D\Gamma}{Dt} = 0,
\end{equation}
where $\Gamma = \oint_C \mathbf{u} \cdot d\mathbf{l}$ is circulation around a material curve $C$. The Helmholtz theorems follow: vortex lines are material lines, vortex tubes cannot end in the fluid, and vortex strength is conserved.

\begin{figure}[h]
    \centering
    \includegraphics[width=1.0\columnwidth]{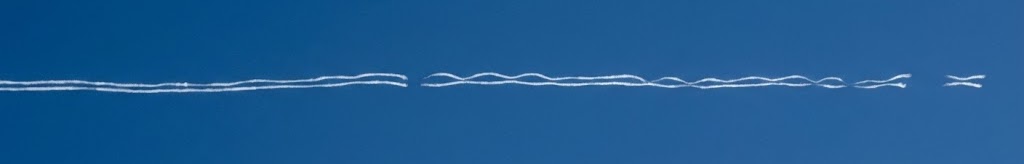}
    \caption{Aircraft contrails exhibiting Crow instability. The sinusoidal deformation of the counter-rotating vortex pair is an inviscid phenomenon driven by mutual induction. However, the eventual pinching, vortex reconnection, and final dissipation require viscosity. In an inviscid world, these vortices would deform and evolve, but persist indefinitely.}
    \label{fig:contrail2}
\end{figure}

Vorticity becomes ``frozen" into the fluid. In our world, viscosity gradually erases vortical structures. Without it, every vortex persists indefinitely, from a stirred whirlpool to aircraft wingtip vortices (Figure \ref{fig:contrail2}), which in our viscous world undergo Crow instability and eventually dissipate, but would deform and persist forever in inviscid flow. At larger scales, oceans and atmosphere would retain every eddy ever generated and result in a chaotic soup of all disturbances throughout history, never fading.

\subsection{Turbulence without viscosity}

Turbulence can exist without viscosity - but its birth, life, and death are fundamentally transformed.

\textbf{Birth}. Viscous turbulence has many origins: boundary layers where no-slip creates shear, free shear layers where streams of different velocities meet, jets injecting momentum into quiescent fluid. What unites them is that viscosity enables the velocity gradients that, beyond a critical Reynolds number, amplify into chaos. Remove viscosity, and these pathways close. Inviscid boundaries enforce only no-penetration ($\mathbf{u} \cdot \mathbf{n} = 0$) - fluid slips freely, no gradient forms. Turbulence must find other origins: baroclinic torque, non-conservative body forces, flow instabilities, or vorticity prescribed at domain
boundaries (e.g. when a jet entering a domain).

\textbf{Life}. Once born, turbulence in both worlds follows Richardson's cascade - large eddies break into smaller ones. The difference is where it ends. Viscous turbulence meets a floor at the Kolmogorov scale: \begin{equation} \eta = \left(\frac{\nu^3}{\epsilon}\right)^{1/4}, \end{equation} where $\nu$ is kinematic viscosity and $\epsilon$ is dissipation rate. Below $\mu$, viscosity halts the cascade and converts motion to heat. Inviscid flow knows no such floor - the cascade continues without limit.

\textbf{Death}. Viscous turbulence dies as heat. Inviscid turbulence has no such exit - the Euler equations contain no temperature. So where does the energy go? Nowhere. Onsager showed that when velocity becomes sufficiently irregular, energy conservation simply fails\cite{onsager1949statistical}. Energy vanishes from the equations, not transformed, just gone. 

\textbf{Mixing}. One of the most familiar manifestations of turbulence in daily life is mixing. But stirring alone is not enough. When we stir cream into coffee, the turbulent motion stretches and folds fluid elements, drawing them into ever-finer filaments. Yet true mixing - where cream and coffee become one uniform fluid - requires molecular diffusion to act across those thinned interfaces. Without diffusion, the filaments would grow infinitely fine but never blend; the two fluids would remain distinct, just impossibly interleaved. Turbulence accelerates mixing not by mixing itself, but by creating the fine-scale structure that lets diffusion finish the job.

\subsection{D'Alembert's paradox}

In steady, irrotational, inviscid, incompressible flow around a body, there is no force - no drag, no lift, nothing. The pressure distribution wraps symmetrically around the object, pushing equally fore and aft. A fish would glide without effort; an airplane would feel no resistance. This is D'Alembert's paradox, and it troubled physicists for over a century. In real flows, drag arises from boundary layer friction and flow separation - both viscous phenomena. While escapes exist, for instance, unsteady motion introduces added mass, circulation produces lift, compressibility creates wave drag, but in the purest inviscid case, the paradox holds. Our everyday intuition that moving through fluid requires effort rests entirely on viscosity.

\section{Day-to-Day Life}

The absence of viscosity would transform mundane activities into hazards and curiosities. This section examines phenomena encountered in everyday experience - from rainfall to drinking water - where viscous dissipation currently operates invisibly but essentially.

\subsection{Lethal rain}

Raindrops normally reach a terminal velocity $u_t$ when gravitational force balances aerodynamic drag:
\begin{equation}
mg = \frac{1}{2} \rho_a C_D A u_t^2,
\end{equation}
where $m$ is drop mass, $g = 9.8\,\mathrm{m/s^2}$ is gravitational acceleration, $\rho_a \approx 1.2\,\mathrm{kg/m^3}$ is air density, $C_D \approx 0.5$ is drag coefficient, and $A$ is frontal area. For a typical raindrop ($d \approx 2\,\mathrm{mm}$), terminal velocity is $u_t \approx 6$-$9\,\mathrm{m/s}$.

Without viscosity, drag vanishes entirely. Raindrops accelerate continuously under gravity. Falling from cloud base at $h \approx 2\,\mathrm{km}$, reaching the speed of $u_{\mathrm{impact}} = \sqrt{2gh} \approx 200\,\mathrm{m/s}$. This exceeds the muzzle velocity of some firearms. A gentle spring shower becomes a barrage of liquid projectiles: umbrellas would be ineffective, and shelter would be mandatory for survival!

\subsection{Meteors and atmospheric entry}

Small meteoroids (mm to cm scale) typically burn up in the atmosphere due to aerodynamic heating. The intense drag decelerates them while frictional heating ablates the material.

Without viscosity, there will be no drag, no deceleration, and no heating from viscous dissipation. Small debris that currently vaporizes harmlessly at $80$-$120\,\mathrm{km}$ altitude would instead reach the surface at cosmic velocities of $v \sim 10,000$-$70,000\,\mathrm{m/s}$ $\approx 40,000$-$250,000\,\mathrm{km/hr}$!

Earth intercepts roughly $10^7$-$10^8\,\mathrm{kg}$ of meteoritic material annually. Currently, most burns up. In an inviscid atmosphere, this mass arrives intact as a continuous bombardment of hypervelocity projectiles. To picture such a world, see the meteor shower scene in the movie \emph{Armageddon} (1998)\cite{armageddon1998}, except that's going to happen non-stop, every day and night!

\subsection{A very loud world}

Sound waves are pressure oscillations. In viscous fluids, acoustic energy dissipates through  (i) viscous absorption (shear stresses between fluid layers), (ii) Thermal conduction (heat flow between compressions and rarefactions) and (iii) Molecular relaxation (energy transfer to internal molecular modes). In dry air, viscous dissipation accounts for more than 50\% of the sound dissipation. The classical absorption coefficient for sound is:
\begin{equation}
\alpha = \frac{\omega^2}{2\rho c^3}\left[\frac{4}{3}\mu + \mu_b + \kappa\left(\frac{1}{c_v} - \frac{1}{c_p}\right)\right],
\end{equation}
where $\omega$ is angular frequency, $c$ is sound speed, $\mu$ is dynamic viscosity, $\mu_b$ is bulk viscosity, $\kappa$ is thermal conductivity, and $c_v$, $c_p$ are specific heats. Without viscosity ($\mu = \mu_b = 0$), the viscous contribution vanishes. However, geometric spreading ($\propto 1/r^2$ for intensity), thermal conduction, and molecular relaxation still operate. Sound remains finite but decays more slowly. What happens in practice is that distant sounds remain audible longer. Traffic noise, construction, conversations, all would carry farther, privacy would diminish, cities would be cacophonous, and Acoustic design of buildings would become critical.

\subsection{Water refuses to stay in containers}

\subsubsection{The Rollin film}

Superfluid helium-4 (below $T_{\lambda} = 2.17\,\mathrm{K}$) exhibits a remarkable behavior: it spontaneously creeps up container walls as a thin film (the Rollin film) and drains out. This occurs because the zero-viscosity superfluid component flows without resistance along any available surface.

If water had zero viscosity, a similar phenomenon would occur. Water would climb container walls via surface tension and drain over the rim. A glass of water left on a table would slowly empty itself, pooling underneath.

The timescale depends on film thickness and surface properties, but the outcome is definite: {open containers cannot hold inviscid liquids indefinitely}.

\subsubsection{Leakage through the tiniest cracks}

In viscous flow, two mechanisms resist water leakage through small pores: (i) {Surface tension:} creates a capillary pressure barrier, and (ii) {Viscosity:} limits flow rate even if pressure overcomes surface tension. 

The capillary pressure required to force liquid through a pore is given by 
\begin{equation}
\Delta P_{\mathrm{cap}} = \frac{2\sigma\cos\theta}{r},
\end{equation}
where $\sigma$ is surface tension coefficient, $\theta$ is contact angle between liquid and surface, and $r$ is pore radius. For water on glass ($\theta \approx 0^{\circ}$, $\sigma \approx 0.073\,\mathrm{N/m}$), a $1\,\mu\mathrm{m}$ pore requires $\Delta P_{\mathrm{cap}} \approx 1.5\times 10^5\,\mathrm{Pa}$ (about $1.5\,\mathrm{atm}$) to initiate flow.

Once flow begins, the volume flow rate through a channel follows Poiseuille's law
\begin{equation}
Q = \frac{\pi r^4 \Delta P}{8\mu L},
\end{equation}
where $\mu$ is viscosity and $L$ is channel length. For example, honey has viscosity $\mu \approx 2$--$10\,\mathrm{Pa \cdot s}$, roughly $10^3$--$10^4$ times that of water ($\mu_{\mathrm{water}} \approx 10^{-3}\,\mathrm{Pa \cdot s}$). Both honey and water pass through a mesh if pore size exceeds the capillary threshold, but the Poiseuille flow rate scales as $Q \propto 1/\mu$. Hence honey flows $10^3$--$10^4$ times slower through the same mesh. 

Without viscosity ($\mu \to 0$), this distinction vanishes: honey pours like water. In other words, without viscosity the Poiseuille flow brake is gone and the flow will be governed by Bernoulli's equation $u = \sqrt{2gh}$, where $h$ is the water depth. Any crack satisfying $\Delta P > \Delta P_{\mathrm{cap}}$ allows water to exit at velocities comparable to flow through large openings. For example, consider a hairline crack at the base of a dam. In our viscous world, Poiseuille scaling limits discharge to a trickle. Without viscosity, water exits the crack at the same speed as through the spillway. For $h = 50\,\mathrm{m}$: $u \approx 31\,\mathrm{m/s}$ through a microscopic crack.

\subsection{Swimming}

In normal swimming, most of the propulsive force comes from viscosity-dependent mechanisms. When a swimmer pulls through the water, the hands and forearms generate both drag and lift forces that push the swimmer forward. The drag arises mostly from pressure differences created by viscous flow separation; the lift arises from circulation around the hand acting as a crude hydrofoil. Additionally, when the hand accelerates water, the reaction force (added mass effect) contributes to thrust. However, added mass typically accounts for a smaller fraction of the total propulsive force (about 20\%), but the dominant contribution comes from viscous drag and lift. Without viscosity, a swimmer's stroke would feel effortless: the hand would slip through the water with almost no resistance. But therein lies the problem: no resistance means no thrust. Only the added mass mechanism survives in inviscid flow, and this requires rapid accelerations rather than steady pulling motions. A swimmer might retain perhaps $20$-$30\%$ of their normal propulsive capability, making acceleration from rest extremely difficult.

Once moving, however, the situation reverses dramatically. The resistance a swimmer faces consists of three components: skin friction drag (viscous shear at the body surface), form or pressure drag (from flow separation behind the body), and wave drag (energy lost to surface waves). At typical swimming speeds ($1$-$2 \, \mathrm{m/s}$), skin friction contributes roughly $5$-$15\%$ of total drag, while pressure drag and wave drag dominate. Wave drag depends on the Froude number $Fr = u/\sqrt{gL}$ (where $u$ is speed, $g$ is gravity, $L$ is body length) and can reach $35$-$60\%$ of total drag at racing speeds near the surface. Both skin friction and form drag arise from viscosity - only wave drag persists in inviscid flow. At low speeds, where wave drag is negligible, an inviscid swimmer would glide almost without deceleration. At higher speeds near the surface, roughly $40$-$60\%$ of the original resistance remains as wave drag. The net effect: once momentum is gained (a difficult task), the swimmer coasts far longer than in the viscous world.

Fish and other aquatic animals, having evolved over hundreds of millions of years in viscous water, would face the same paradox. Their undulatory and oscillatory propulsion relies on generating vortices and pressure differences - mechanisms rooted in viscosity. As Purcell observed in his celebrated essay \emph{``Life at Low Reynolds Number''} (1977)\cite{purcell1977lowreynolds}, locomotion strategies depend critically on the hydrodynamic regime. While Purcell examined the opposite extreme (microscopic life dominated by viscosity) inviscid flow presents equally fundamental challenges. Evolution would need to discover entirely new propulsion strategies, perhaps based on jet propulsion (expelling water rearward) or exploiting added mass through impulsive, jerky motions rather than graceful strokes.

\section{Flight in an Inviscid World}

\subsection{The death of conventional lift}

In our viscous atmosphere, flight is elegantly simple. As an aircraft accelerates along a runway, the relative wind over the wing, combined with viscosity, generates vorticity within a thin boundary layer. This vorticity establishes circulation $\Gamma$ around the airfoil. By the Kutta-Joukowski theorem, the lift per unit span is
\begin{equation}
L' = \rho U \Gamma,
\end{equation}
where $\rho$ is air density, $U$ is freestream velocity, and $\Gamma$ is circulation. The Kutta condition which is satisfied only because viscosity acts at the sharp trailing edge, ensures a unique, nonzero $\Gamma$. The aircraft becomes airborne.

In an inviscid world, this mechanism fails entirely. Without viscosity, no boundary layer forms, no vorticity is generated at the surface, and the Kutta condition cannot be enforced. The result is d'Alembert's paradox: potential flow closes smoothly around the wing, yielding zero lift and zero drag. An aircraft could accelerate to $300\,\mathrm{m/s}$ on the runway and feel nothing from the air: no resistance, but also no lift. Traditional wings become decorative appendages.

To achieve lift without viscosity, we must impose circulation artificially. One approach involves distributed wall jets around the wing surface, injecting momentum to force circulation. If $\Gamma$ is maintained and the aircraft has forward speed $U$, Kutta-Joukowski still applies and lift emerges around any bluff shape, not just streamlined airfoils. The aesthetic consequence is striking: aircraft need not be sleek. A flying brick would work as well as a wing, provided the circulation is maintained. Alternatively, lift can be achieved by vertical thrust (e.g. rocket engines), but that's likely a much more expensive way of maintaining lift throughout the flight.

Cruising, on the other hand, would be efficient. With no viscous drag, an aircraft in steady flight requires no thrust to maintain forward motion, only enough power to sustain/adjust circulation via its wall jets or rocket engines.

Landing presents another stark challenge. In a viscous world, descent dissipates most of the kinetic energy of the aircraft through viscous drag. In other words, the atmosphere acts as a gentle brake. In an inviscid world, d'Alembert's paradox ensures zero drag. An aircraft retains every Joule of kinetic energy acquired during acceleration. To decelerate from cruising speed to touchdown, reverse thrust equal in magnitude to that used for takeoff must be applied. The fuel implications are severe. A flight requires fuel for: (1) acceleration to cruising speed, (2) circulation modifications during cruise, and (3) deceleration to landing speed. Items (1) and (3) are roughly symmetric: whatever energy was put in must be actively removed. Aircraft of the inviscid world, if they run out of fuel, cannot land and are doomed!

\subsection{Propulsion without airfoils}

The problem extends to propulsion. Propellers and jet engine compressor blades are rotating assemblies of airfoils. Without viscosity to generate circulation on these blades, compressors cannot compress, turbines cannot extract work, and propellers cannot generate thrust in the conventional sense. A turbofan engine becomes an elaborate paperweight! 

What propulsion options remain? Rockets work: they expel mass and generate thrust by Newton's third law, independent of the surrounding fluid's properties. Ducted fans with internal flow acceleration (relying on pressure differences rather than blade lift) could provide some thrust, though less efficiently. The inviscid aircraft would resemble a rocket plane more than a 747: pure reaction propulsion for forward motion, plus whatever circulation-imposing system keeps it aloft.

\subsection{Parachutes and emergency descent}

Parachutes become useless ornaments. A parachute generates drag through pressure differences created by flow separation and wake formation - viscous phenomena. In potential flow, the air closes smoothly behind the canopy. A skydiver with a fully deployed parachute would hit the ground at the same speed as without one: $v = \sqrt{2gh}$, where $g \approx 9.8\,\mathrm{m/s^2}$ is gravitational acceleration and $h$ is drop height. From $h = 4000\,\mathrm{m}$ (a typical skydiving altitude), impact velocity would be approximately $280\,\mathrm{m/s}$ - supersonic and invariably fatal.

Emergency procedures would require onboard retro-rockets or explosive deceleration systems. Ejection seats would need integrated rocket packs to arrest descent. The romance of floating gently under a canopy would never have entered human imagination.

\subsection{Vortex sheets and atmospheric hazards}

Without viscous diffusion, vorticity cannot spread (i.e. diffuse) or decay. Vortex sheets (surfaces of discontinuous velocity) would persist indefinitely with infinite shear rates. Every disturbance in the atmosphere, from a passing truck to a distant thunderstorm, would leave behind permanent velocity discontinuities.

\begin{figure}[h]
    \centering
    \includegraphics[width=0.8\columnwidth]{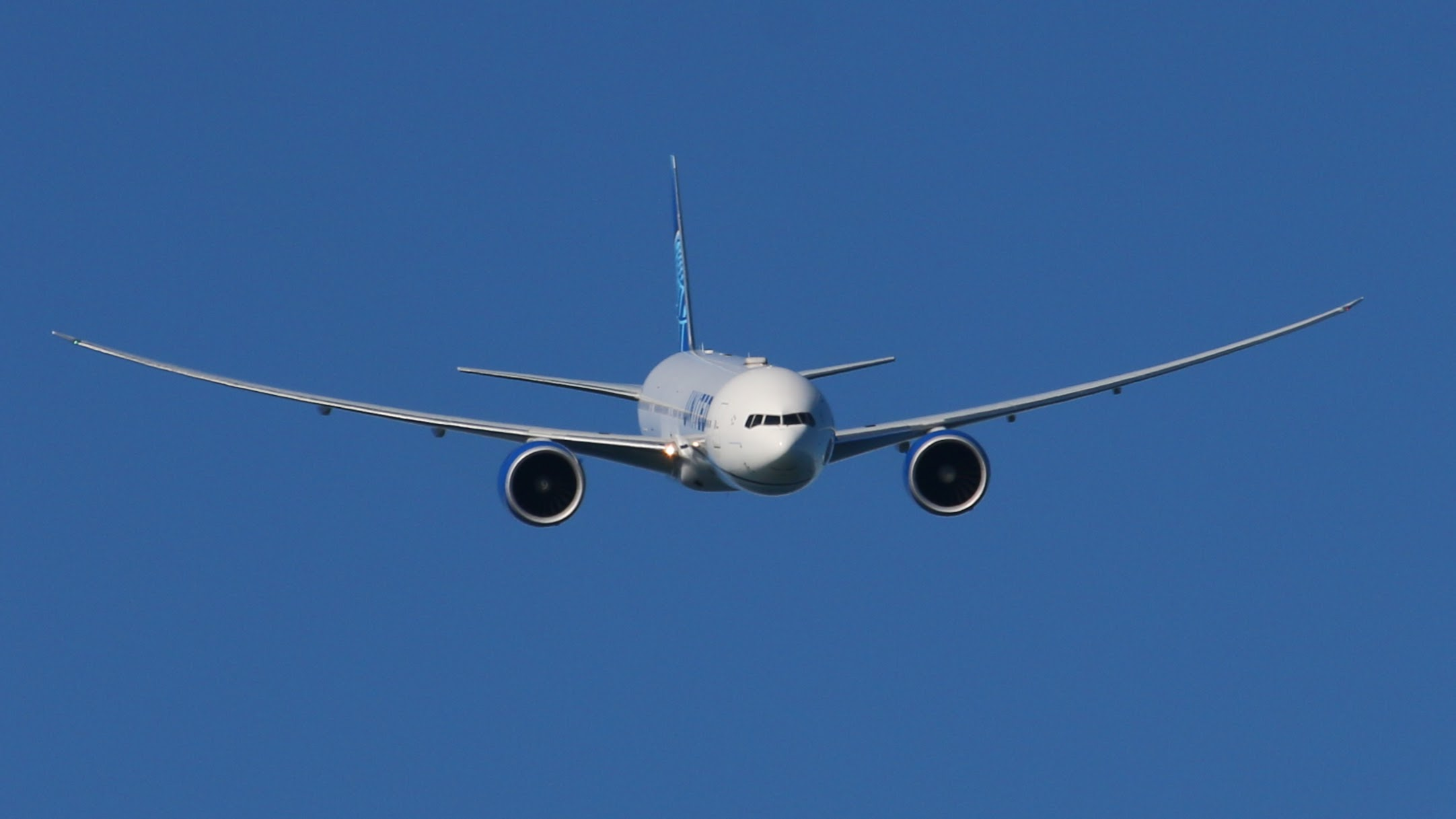}
    \caption{A Boeing 777 during a pitch-up recovery at approximately 2$g$, showing visible wing flex near the upper end of its design load envelope. In an inviscid atmosphere, vortex-sheet encounters could impose loads potentially orders of magnitude beyond this limit, making even routine flight structurally untenable (\textit{San Francisco Airshow, 10/12/2025}).}
    \label{fig:777}
\end{figure}

For aircraft, this creates a nightmare scenario. Encountering a vortex sheet means instantaneous, extreme wind shear. Structural loads scale with velocity change; a sudden $\Delta U = 50\,\mathrm{m/s}$ across a vortex sheet would impose accelerations of hundreds of $g$ on the airframe, far exceeding the $-1g$ to $+2.5g$ design limits of today's commercial aircraft (see figure \ref{fig:777}). Catastrophic structural failure would be routine. Flight paths would need to thread through an invisible, permanent obstacle course of frozen vorticity.

\subsection{Biological flight: an evolutionary dead end}

Birds and insects face identical constraints. In our viscous world, flapping wings generate lift through the same viscosity-dependent mechanism: unsteady circulation established via the Kutta condition at the trailing edge. A hummingbird's wing, beating at $80\,\mathrm{Hz}$, relies on viscosity to create the bound vorticity that supports its $4$-gram body.

In an inviscid world, biological flight is impossible. No evolutionary pathway leads to wings that work without viscosity. Flying animals would perhaps never have evolved (or evolved in an entirely different way). The skies would be silent: no birds, no bats, no insects on the wing. Pollination would be restricted ground-based creatures (not even wind can help). Ecosystems would be unrecognizable.

The absence of flying insects has cascading consequences. Approximately $75\%$ of flowering plants depend on animal pollinators; without flying insects, terrestrial ecosystems would collapse into something far simpler. The ``inviscid Earth'' would be a quieter, emptier, less colorful place.

In a world without viscosity, aviation reduces to rocketry. Every aircraft becomes a flying fuel tank: thrust to accelerate, active systems to generate lift, thrust again to decelerate. The grace of soaring birds, the efficiency of gliders, the elegance of a bird catching a thermal - all impossible. Orville and Wilbur Wright, staring at their flyer at Kitty Hawk, would have watched it skid uselessly down the rail, the wind passing through and around the wings as if they were not there.

Perhaps the deepest irony is that we owe the magic of flight to what engineers usually consider a nuisance: friction. Viscosity is the hidden partner in every wing, every feather, every insect that has ever taken to the air.

\section{Engineering in an Inviscid World}

\subsection{The collapse of rotating machinery}

Perhaps no engineering application depends more critically on viscosity than lubrication. In a journal bearing, a rotating shaft floats on a thin film of oil, typically $10$-$100\,\mu\mathrm{m}$ thick. The shaft does not touch the housing because viscous shear in the lubricant generates pressure that supports the load. This hydrodynamic lubrication, described by the Reynolds equation, requires viscosity as its essential ingredient.

Without viscosity, the oil film cannot support any load. Metal contacts metal directly. Friction coefficients jump from $\sim 0.001$ (hydrodynamic) to $\sim 0.3$--$0.5$ (dry metal-on-metal). Surfaces weld microscopically under pressure and shear - a phenomenon called galling or seizure. Every rotating machine - turbines, generators, vehicle engines, and industrial motors - would grind to a halt within seconds of startup. The industrial revolution, built on spinning shafts, would be impossible.

\subsection{Hydraulic systems: power transmission fails}

Hydraulic systems transmit power through pressurized fluid. A pump pressurizes oil; that pressure actuates pistons, lifts loads, and drives machinery. The system works because viscous fluid resists leaking past seals and through clearances.

In an inviscid world, fluid bypasses every gap instantaneously. Pistons cannot build pressure - oil flows freely around them. Hydraulic excavators, aircraft control surfaces, automobile brakes, and industrial presses all fail. Construction equipment becomes inoperable. The infrastructure we use to build infrastructure vanishes.

\subsection{Dams and water retention}

Earthfill dams rely on viscosity to limit seepage. Darcy's law governs flow through porous media:
\begin{equation}
Q = \frac{\kappa A \Delta P}{\mu L},
\end{equation}
where $Q$ is volumetric flow rate, $\kappa$ is permeability, $A$ is cross-sectional area, $\Delta P$ is pressure difference, $\mu$ is dynamic viscosity, and $L$ is seepage path length. Similar to what we discussed for Poiseuille flow, if $\mu \to 0$, this mechanism of flow control fails.

Surface tension cannot help: soil particles are typically hydrophilic, so water wets them readily. Capillary forces actually promote infiltration rather than resist it. Only concrete gravity dams - relying on impermeability rather than viscous resistance - would function, and even these would require perfect seals at every joint.

\subsection{Dampers and Vibration Control}

Conventional shock absorbers dissipate energy by forcing fluid through orifices, producing a damping force proportional to velocity: $F_d = c\dot{x}$. This yields smooth, exponential decay of oscillations. Without viscosity, this mechanism vanishes.

Other dissipation mechanisms persist, however. Coulomb (dry friction) damping arises from sliding surfaces, with force independent of speed. Hysteretic damping results from internal friction within deforming materials. A car hitting a bump would not oscillate forever because rubber bushings, metal flexure, and joint friction still extract energy. But the decay would be linear rather than exponential, and the ride harsh and jerky. The elegant tunability of viscous dampers would be lost. Alternatives exist: friction-based dampers (which wear and behave nonlinearly) or active control systems (which require power and complex electronics). Neither matches the simplicity and reliability of passive viscous devices.

\subsection{Flow Control and Valves}

Valves regulate flow by forcing fluid through a restriction. In viscous flow, this creates a permanent pressure drop because energy is dissipated as fluid navigates the constricted path, reducing both flow rate and velocity in a controlled manner.

In inviscid flow, Bernoulli's equation governs. Partially closing a valve reduces the open area, which reduces volume flow rate by continuity ($Q = Av$). However, the fluid velocity through the opening remains unchanged, set entirely by the upstream-downstream pressure difference. A tap connected to a pressurized main would produce a high-speed jet through any opening, no matter how small. Pressure regulators, which function by dissipating energy through variable restrictions, would fail entirely because there is no dissipation mechanism to exploit. Gravity-fed pipelines pose an even greater hazard. With zero resistance, fluid accelerates freely downhill. Velocities can exceed the local sound speed in the liquid, and pressure can drop below vapor pressure, triggering explosive cavitation. Preventing this would require continuous throttling along the pipeline which is precisely the capability that inviscid flow denies. Flow control reduces to binary on-off operation, with no means to moderate velocity or dissipate energy gradually.

\subsection{Summary}

Many everyday technologies exploit viscosity in subtle ways. Paint adheres partly through viscous flow that wets surfaces and levels before drying; without it, coatings run off fast, and only baked-on or electrostatic finishes would work. HVAC systems distribute air through ducts sized to create specific pressure drops; without viscous resistance, air flows through the shortest path regardless of damper settings, making climate control impossible. The common thread is that viscosity provides controllable resistance - the ability to dissipate energy gradually, support loads through fluid films, and meter flows predictably. Engineering in an inviscid world would require complete reinvention: magnetic or rolling-element bearings instead of hydrodynamic ones, electromechanical actuators instead of hydraulics, friction-based or active vibration control, and positive shutoff valves with perfect seals. Without viscosity, engineering becomes a binary world of zero resistance or solid contact, with nothing in between.

\section{Biology in an Inviscid World}

The cardiovascular system is perhaps the most viscosity-dependent structure in animal biology. Blood viscosity - approximately $\mu \approx 3$-$4 \, \mathrm{mPa \cdot s}$, about three times that of water - provides the resistance that allows hearts to generate and maintain pressure gradients. The smallest arterioles, with radii of $10$-$50 \, \mu\mathrm{m}$, create the dominant resistance that sets blood pressure. If viscosity vanishes, so does vascular resistance. The heart, evolved to pump against $\sim 100 \, \mathrm{mmHg}$, would suddenly face near-zero load. Blood would accelerate uncontrollably with each heartbeat, and organ perfusion (the controlled delivery of blood to tissues) would become impossible.

Wounds compound the problem. In our viscous world, small injuries bleed slowly because resistance scales inversely with opening size, and clotting factors accumulate at wound sites to form stable plugs. Without viscosity, blood accelerates through even microscopic openings while clotting factors wash away before aggregating. A paper cut could produce catastrophic bleeding. Even intact circulation fails at the exchange level: oxygen delivery requires slow capillary flow (roughly $0.5$-$1 \, \mathrm{mm/s}$) giving red blood cells $1$-$2 \, \mathrm{s}$ to exchange gases with tissue. Without viscous resistance, transit times drop to milliseconds, and tissues become hypoxic despite adequate blood supply.

Plants face analogous challenges. Water ascends through xylem vessels via cohesion-tension: transpiration at leaves pulls a continuous water column upward, and viscous resistance in narrow conduits stabilizes this flow. Without it, the metastable water column would fragment or oscillate violently. Tall trees, some exceeding $100 \, \mathrm{m}$, could not maintain stable hydraulic columns, and forests would not exist. The respiratory system tells a similar story: mucus, a high-viscosity gel, traps inhaled pathogens while cilia beat in coordinated waves to push it upward. Without viscosity, mucus cannot trap particles, cilia lose their purchase, and pathogens accumulate in the lungs unopposed.

At larger scales, synovial joints rely on hydrodynamic lubrication: viscous fluid films that support load so cartilage never touches directly. Without viscosity, friction coefficients jump from $\sim 0.001$ to $\sim 0.1$-$0.3$, and joint surfaces grind with each step. At the opposite extreme, microorganisms cannot move at all. Bacteria and sperm swim by pushing against viscous drag; at cellular scales ($Re \sim 10^{-4}$), there is nothing else to push against. In inviscid fluid, flagella spin freely without propelling cells, sperm become immotile, and fertilization fails (unless perhaps if they evolve to use recoil-based propulsion (ejecting mass) which doesn't rely on viscosity).

Biology is built on viscous flow at every scale. In an inviscid world, complex life - multicellular animals, vascular plants - could never evolve. The viscosity of water is not an inconvenience; it is the foundation on which life stands.

\section{Environment in an Inviscid World}

\subsection{The atmospheric boundary layer vanishes}

The atmospheric boundary layer, the lowest $1$--$2\,\mathrm{km}$ of the atmosphere where surface friction slows winds, depends entirely on viscosity. In our world, wind stress at the surface transfers momentum downward, creating a velocity gradient from geostrophic winds aloft (a balance between Coriolis and pressure gradient, resulting in often $30$-$50\,\mathrm{m/s}$ in jet streams) to near-calm conditions at ground level. This gradient is maintained by turbulent (effective) viscosity.

Without viscosity, boundary layers cannot form. Air slips freely over land and ocean with negligible drag. Surface winds approach free-atmospheric velocities. A typical mid-latitude location might experience sustained winds of $30$-$50\,\mathrm{m/s}$ (hurricane force) as the new normal. The gentle breezes we associate with pleasant weather would not exist; every day would feel like standing in a wind tunnel.

\subsection{Storms that never die}

Hurricanes and cyclones currently dissipate through two primary mechanisms: surface friction over land and ocean, and turbulent mixing that entrains dry air and disrupts the warm-core structure. Viscous dissipation converts kinetic energy to heat at a rate proportional to $\mu$; with $\mu \to 0$, this pathway closes.

A hurricane making landfall would not weaken from friction. It would lose energy only through thermodynamic limits (i.e. running out of warm ocean to feed it) or by radiating energy as waves. Once formed, tropical cyclones could persist for months or years, circumnavigating the globe multiple times. Historical storms like Hurricane John (1994)\cite{hurricanejohn1994}, which lasted $31$ days, would become the norm rather than the exception.

The same logic applies to mid-latitude cyclones, anticyclones, and blocking patterns. Weather systems would become quasi-permanent features. Climate zones would be defined not by latitude alone but by which long-lived vortex happened to dominate a region. Jupiter's Great Red Spot, a storm persisting for centuries, offers a glimpse of this regime.

\subsection{Groundwater and rivers}

Groundwater flow through porous aquifers follows Darcy's law: flow rate is proportional to pressure gradient and inversely proportional to viscosity. This inverse dependence means viscosity acts as nature's throttle, slowing water's passage through soil and rock to timescales of years or millennia. Aquifers recharge gradually and release water steadily, sustaining wells, springs, and river baseflow between storms. With zero viscosity, any pressure gradient produces unbounded flow. Aquifers would drain in hours rather than millennia; groundwater storage, humanity's largest freshwater reserve, would empty into the oceans after a single storm.

Rivers suffer a similar fate. Normal flow balances gravitational driving force against frictional resistance, keeping velocities moderate - perhaps $1 \, \mathrm{m/s}$ for a gentle stream. Without viscous resistance, that same stream might accelerate to $10 \, \mathrm{m/s}$ or more. Rivers would become violent, erosive torrents. Flood waves would propagate downstream at full strength rather than attenuating, and riverbeds would reorganize continuously under high-speed assault. Between floods, rivers would run dry: no slow groundwater seepage to sustain them. The stable, life-supporting hydrology we know would be replaced by cycles of catastrophic deluge and desiccation.

\subsection{Magma and volcanism}

Magma viscosity spans an enormous range, from $10^2 \, \mathrm{Pa \cdot s}$ for fluid basalt to $10^{12} \, \mathrm{Pa \cdot s}$ for stiff rhyolite, and this variation controls eruption style. High-viscosity magma traps gases, building pressure until explosive release; low-viscosity magma lets gases escape gently. The viscosity of molten rock determines whether a volcano oozes or detonates.

With zero viscosity, magma would flow like water. Conduit resistance would vanish, and molten rock would ascend from chambers at extraordinary rates. Eruptions would become catastrophic gushers, neither the slow lava flows of Hawaii nor the explosive blasts of Vesuvius, but something far worse. Lava would travel vast distances before cooling, advancing at kilometers per minute rather than meters per hour. Deeper still, mantle convection, currently throttled by rock viscosity of $10^{19}$-$10^{21} \, \mathrm{Pa \cdot s}$ to timescales of millions of years, would accelerate dramatically. Plate tectonics would transform from a slow geological dance into a rapid, violent upheaval. The solid Earth we stand on is stable only because its interior resists flow.

Viscosity is Earth's thermostat for mechanical energy: it damps storms, calms seas, slows rivers, and throttles magma - without it, the planet accumulates motion without limit and becomes uninhabitable.

\section{Conclusion}

This thought experiment reveals viscosity as a hidden architect of the physical world. By removing a single material property, internal fluid friction, we have shown that flight, engineering, biology, and environmental stability all collapse.

The consequences are quantifiable. Lift generation fails because the Kutta condition requires viscous action at trailing edges. Journal bearings cannot support loads when $\mu \to 0$ in the Reynolds equation. Blood circulation becomes uncontrollable as vascular resistance vanishes per Hagen-Poiseuille ($R \propto \mu$). Aquifers drain instantly under Darcy's law ($Q \propto 1/\mu$). Atmospheric boundary layers disappear, exposing surfaces to jet-stream velocities of $30$-$50\,\mathrm{m/s}$ continuously.

The pedagogical value is clear: viscosity provides resistance, damping, and control, the ability to dissipate energy gradually rather than catastrophically. Evolution, engineering, and Earth's climate all exploit this property at every scale, from cytoplasmic streaming to hurricane decay. The inviscid limit is not merely an idealization; it is a window into why our viscous world functions as it does.

Future work might extend this analysis to non-Newtonian fluids, where viscosity varies with shear rate, or to quantum fluids where superfluidity offers real-world glimpses of zero-viscosity behavior. Computational studies comparing Navier-Stokes and Euler solutions for biological and environmental flows could quantify exactly where viscosity matters most. Such investigations would deepen our appreciation for a property that, despite its ubiquity, remains underappreciated in scientific education.



\bibliographystyle{aipnum4-2}
\bibliography{refs}

\end{document}